\input{aipcheck}

\documentclass[
    ,final            % use final for the camera ready runs
  ]
  {aipproc}

\layoutstyle{8x11double}

\begin{document}
\def\be{\begin{equation}}
\def\ee{\end{equation}}
\def\bea{\begin{eqnarray}}
\def\eea{\end{eqnarray}}
\def\tr{{\rm tr}\, }
\def\nn{\nonumber \\}
\def\e{{\rm e}}

\title{Viable f(R) gravity and future cosmological evolution}


\author{Diego S\'aez-G\'omez}{
  address={Fisika Teorikoaren eta Zientziaren Historia Saila, Zientzia eta Teknologia Fakultatea, Euskal Herriko Unibertsitatea, 644 Posta Kutxatila, 48080 Bilbao, Spain, EU},
  email={diego.saez@ehu.es},
  %thanks={This work was commissioned by the AIP}
}

% \copyrightholder{Acoustical Scociety of America}
%\copyrightyear  {2012}

\begin{abstract}
One of the so-called viable modified gravities is analyzed. This kind of gravity theories are characterized  by a  well behavior at local scales, where General Relativity is recovered, while the modified terms become important at the cosmological level, where the late-time accelerating era is reproduced, and even the inflationary phase. In the present work,  the future cosmological evolution for one of these models is studied. A transition to  the phantom phase is observed. Furthermore, the scalar-tensor equivalence of $f(R)$ gravity is also considered, which  provides important information concerning this kind of models.
\end{abstract}

%\date{\today}

\maketitle

%%%%%%%%%%%%%%%%%%%%%%%%
\section{Introduction}
%%%%%%%%%%%%%%%%%%%%%%%%
As a consequence of the discovery of the accelerating expansion of the universe, a lot of dark energy candidates have been proposed along recent years. Modified gravity theories has become one of the most popular candidate, since it provides a way to understand the problem of the dark energy and the possibilities to reconstruct a gravitational theory capable to reproduce late time acceleration, and even the inflationary phase (for a review see Ref.~\cite{0601213}). In this sense, $f(R)$ theories are the simplest ones, since they consist in a generalization of the Hilbert-Einstein action to a more complex function of the Ricci scalar,
\be
S=\int d^4x \sqrt{-g}\left[f(R)+2\kappa^2\mathcal{L}_m\right] \ .
\label{1.1}
\ee
Here the coupling constant is as usual given by $\kappa^2=8\pi G$, while $\mathcal{L}_m$ is the matter Lagrangian. The field equations corresponding to the action (\ref{1.1}) are obtained by the variation of this action with respect to the metric tensor $g_{\mu\nu}$, what yields
\[
 R_{\mu\nu} f_R(R)- \frac{1}{2} g_{\mu\nu} f(R) + g_{\mu\nu}  \nabla_{\alpha}\nabla^{\alpha} f_{R}(R) -  \nabla_{\mu} \nabla_{\nu}f_{R}(R)=
\]
\be
= \kappa^2T^{(m)}_{\mu\nu}\ .
\label{1.2}
\ee
Here the subscript $_{R}$ denotes derivatives with respect to $R$. Here we are interested to study flat Friedmann-Lema\^itre-Robertson-Walker (FLRW) metrics, 
\be
ds^2=-dt^2+a(t)^2\sum_{i=1}^3dx^{i2}\ .
\label{metric}
\ee
Hence, modified FLRW equations are obtained through the field equations (\ref{1.2}),   
\[
H^2=\frac{1}{3f_R}\left[\kappa^2 \rho_m +\frac{Rf_R-f}{2}-3H\dot{R}f_{RR}\right]\ , 
\]
\[
-3H^2-2\dot{H}=\frac{1}{f_R}\left[\kappa^2p_m+\dot{R}^2f_{RRR}+2H\dot{R}f_{RR}+ \right.
\]
\be
\left. \ddot{R}f_{RR}+\frac{1}{2}(f-Rf_R)\right]\ ,
\label{1.3}
\ee
where dots denote derivatives with respect to the cosmic time, $H(t)=\dot{a}/a$ is the Hubble parameter, and Ricci scalar is given by $R=6\ (2H^2+\dot{H})$ for the metric (\ref{metric}). Hence, any cosmology can be reproduced by the appropriate action (see Ref.~\cite{0908.1269}). In this manuscript, we focus on the study of a subclass of modified gravities, the  so-called viable $f(R)$ gravities, as they accomplish some indispensable conditions to be considered realistic candidates to describe the universe evolution. This kind of $f(R)$ gravities are usually described by actions of the type $f(R)=R+F(R)$, 
%\be
%f(R)=R+F(R)\ .
%\label{1.4b}
%\ee
which basically represents Hilbert-Einstein action plus an additional term that should affect only at cosmological scales, while at local scales, General Relativity is recovered. Viable gravities are able to satisfy  this constraint, as well as avoids large instabilities in the presence of matter distributions, and the anti-gravity regime, by the appropriate form of the function $F(R)$ (see \cite{f(R)viable1}-\cite{f(R)viableOthers}). Moreover, these theories are capable to reproduce the effects of dark energy, and even inflation. Note that both equations in (\ref{1.3}) are written in such a way that  $F(R)$-terms are on the matter side. We can define an effective energy density  for the extra $F(R)$-terms, so that the first FLRW equation in (\ref{1.3}) can be rewritten for the kind of actions in terms of the cosmological parameters $\Omega_i$,
\[
1=\Omega_{m}+\Omega_{F(R)}\ , 
\]
where,
\[
\Omega_m=\frac{\rho_m}{\frac{3}{\kappa^2}H^2}\ ,  
\]
\be
\Omega_{F(R)}=\frac{1}{3H^2}\left(\frac{RF_R-F}{2}-3H\dot{R}F_{RR}-3H^2F_R\right)\ .
\label{1.4}
\ee
Here, we have assumed $f(R)=R+F(R)$. Then, the first Friedmann equation (\ref{1.3}) takes a simple form, with two fluids contributing to the scale factor dynamics. In addition, the continuity equation $\nabla_{\mu}T^{\mu\nu}=0$ for a perfect fluid with an equation of state (EoS) $p_m=w_m\rho_m$ yields $\dot{\rho}_m+3H(1+w_m)\rho_m=0$. Hence, for a particular $F(R)$,  the corresponding cosmological evolution can be obtained through equations(\ref{1.3}), and the continuity equation.  Then, under these circumstances, the $F(R)$ term contributes to the dynamics of the scale factor, and it is capable to reproduce both accelerating epochs of the universe evolution. Here we will focus on the study of one model of this kind of viable modified gravities, proposed by Hu and Sawicki in Ref.~\cite{f(R)viable1}, and whose action is given by,
\be
F_{HS}(R)=-R_{HS}\frac{c_1(R/R_{HS})^n}{c_2(R/R_{HS})^n+1}\ ,
\label{1.8a}
\ee
where $\{c_1, c_2,n\}$ are free parameters and $R_{HS}=\kappa^2\rho^0_m$. The next sections are devoted to the study of the cosmological evolution for this model, specially during the dark energy epoch, and in the future. We will also study the behavior of their scalar-tensor counterpart, where the phase space is explored. This analysis is based on the study performed in Ref.~\cite{SaezGomez:2012ek}.

%%%%%%%%%%%%%%%%%%%%%%%%%%%%%
\section{Cosmological evolution}
%%%%%%%%%%%%%%%%%%%%%%%%%%%%%
In this section, we explore the cosmological evolution for the Hu-Sawicki model. For convenience, we express the equations (\ref{1.3}) in terms of the redshift $z$ instead of the cosmic time $t$,
$1+z=\frac{a_0}{a(t)}$, where $a_0$ is the value of the scale factor at the present time $t_0$, such that the current epoch corresponds to $z=0$. 
\begin{figure}[h!]
  \resizebox{14pc}{!}{\includegraphics{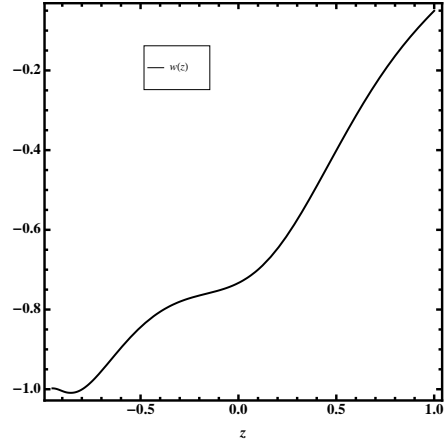}}
  %\resizebox{13.7pc}{!}{\includegraphics{1b}}
\caption{Evolution of the EoS parameter (\ref{2.5}) as a function of the redshift $z$ for the model (\ref{1.8a}) with $n=1$, $c_1=2$ , $c_2=1$, and initial conditions $h(0)=1$ and $h'(0)=0.4$.}
\label{fig1}
\end{figure}
\begin{figure}[h!]
%  \resizebox{13pc}{!}{\includegraphics{1a}}
  \resizebox{14pc}{!}{\includegraphics{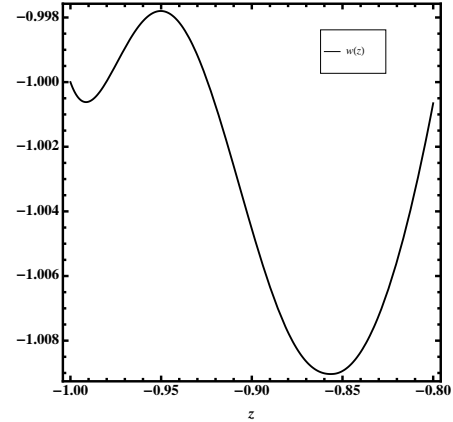}}
\caption{Evolution of the EoS in the interval $-1<z<-0.8$, plotted in more detail, where the EoS crosses clearly the phantom barrier and oscillates.}
\label{fig1a}
\end{figure}

\begin{figure}[h!]
  \resizebox{14pc}{!}{\includegraphics{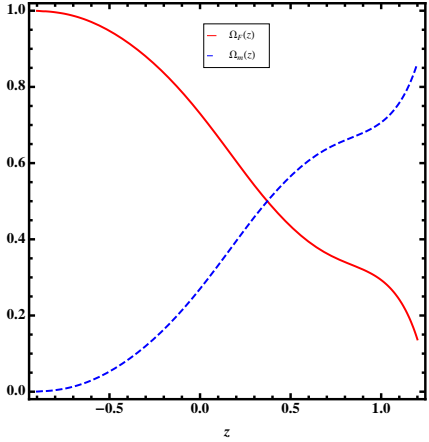}}
  %\resizebox{13.7pc}{!}{\includegraphics{1b}}
\caption{Evolution of the cosmological parameters $\{\Omega_m,\Omega_{F}\}$ defined in (\ref{1.4}) as a function of the redshift $z$ for the Hu-Sawicki model.}
\label{fig2}
\end{figure}

%Then, the time derivative is transformed as $\frac{d}{dt}=-(1+z)H\frac{d}{dz}$, and  the first FLRW equation in (\ref{1.3}) and the continuity equation (\ref{1.4a}) yield,
%\be
%H^2(z)=\frac{1}{3f_R}\left[\kappa^2\rho_m(z)+\frac{R(z)f_R-f}{2}+3(1+z)H^2f_{RR}R'(z)\right]\ ,
%\label{2.2}
%\ee
%\be
%(1+z)\rho_m'(z)-3(1+w_m)\rho_m(z)=0\ ,
%\label{2.3}
%\ee
%where now primes denote derivatives with respect to the redshift. Then, the Ricci scalar can be rewritten as $R=6\left[2H^2(z)-(1+z)H(z)H'(z)\right]$, while the equation (\ref{2.3}) can be easily solved for a constant EoS parameter $w_m$,
%\be
%\rho(z)=\rho_0(1+z)^{3(1+w_m)}\ .
%\label{2.4}
%\ee
Then, we can fit the current values of the cosmological parameters using the observational data \cite{WmapData}, where $H_0=100\, h \ km\ s^{-1}\ Mpc^{-1}$ with $h=0.71\pm 0.03$ and the matter density, $\Omega_m^0=0.27 \pm 0.04$, while the matter fluid is considered pressureless (cold dark matter and baryons), $w_m=0$.  Then, by setting the initial conditions,  the corresponding cosmological evolution can be obtained through the Hubble parameter in terms of the redshift and the future evolution can be explored (with $-1<z<0$). Here, we analyze the evolution of the effective EoS $w_{eff}$, which is defined as
\be
w_{eff}=\frac{p_{F(R)}+p_m}{\rho_{F(R)}+\rho_m}=-1-\frac{2\dot{H}(t)}{3H^2(t)}=-1+\frac{2(1+z)H'(z)}{3H(z)}\ .
\label{2.5}
\ee
For simplicity we redefine the Hubble parameter as $H(z)=H_0\ h(z)$, such that $h(0)=1$, while the initial condition on the first derivative $h'(0)$ can be fixed by assuming that our model mimics $\Lambda$CDM at $z=0$, which basically gives $h'(0)=\frac{\kappa^2}{2H_0^2}\rho_m^0=\frac{3}{2}\Omega_m^0$.
Let's now analyze the $F_{HS}(R)$ model in (\ref{1.8a}),  assuming $c_1=2$ and $c_2=1$, which are dimensionless parameters, and a power of $n=1$. In Fig.~\ref{fig1a}, the evolution of the EoS parameter (\ref{2.5}) is plotted with respect to the redshift for the initial conditions that mimic $\Lambda$CDM model at $z=0$. In spite of the evolution of the EoS parameter is quite similar to the $\Lambda$CDM model for positive redshifts, the universe enters in a phantom phase in the future (negative redshifts), where the EoS parameter turns out less than $-1$, and it oscillates close to $z=-1$ as shown in more detail in Fig.~\ref{fig1a}. In Fig.~\ref{fig2},  the evolution of the cosmological parameters $\{\Omega_m,\Omega_F\}$ with respect to the redshift is plotted, where $\Omega_F$ tends to dominate completely the universe in the future.  
Note that the viable modified model considered here (\ref{1.8a}) produces some oscillations along the cosmological evolution. In general, the transition to the phantom epoch occurs for this class of models, as pointed out in \cite{SaezGomez:2012ek}, which does not imply the existence of a singularity in the future. Nevertheless, a singularity may occur along the whole evolution, since the scalar potential presents a pole for a finite value of the scalar field. However, the singularity may be excluded from the range of interest by a particular choice of the free parameters.

%%%%%%%%%%%%%%%%%%%%%%%%%%%%%
\section{Scalar-tensor counterpart}
\begin{figure}[h!]
  \resizebox{14pc}{!}{\includegraphics{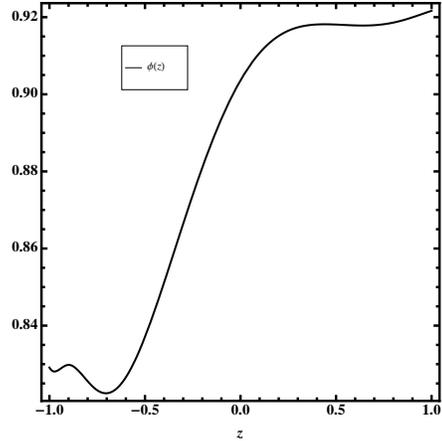}}
  %\resizebox{13.7pc}{!}{\includegraphics{1b}}
\caption{Evolution of the scalar field $\phi(z)$, (\ref{3.7}). As shown, there is a correlation between the evolution of the EoS parameter in Figs.~\ref{fig1}-\ref{fig1a}, and the evolution of the scalar field.}
\label{fig3}
\end{figure}

%%%%%%%%%%%%%%%%%%%%%%%%%%%%%
It is well known that $f(R)$ gravity is equivalent to a kind of Brans-Dicke theory with a null kinetic term, and a scalar potential, (see for example, \cite{MyF(R)1} and references therein),
 \be
S=\int d^4x \sqrt{-g}\left[\phi\ R-V(\phi)+2\kappa^2\mathcal{L}_m\right] \ .
\label{3.1}
\ee
where the scalar field and the potential are related with a particular $f(R)$ action by, 
\be
\phi=f_R(R)\ , \quad V(\phi(R))=f_R(R)\ R-f(R)\ .
\label{3.3}
\ee
Let's reconstruct the corresponding scalar-tensor theory for the particular model considered in the previous section. By the expressions (\ref{3.3}), the corresponding relation $\phi(R)$ for the model   (\ref{1.8a}) is easily obtained,
\be
\phi_{HS}=1+\frac{c_1 c_2 R}{R_{HS}\left(1+\frac{c_2 R}{R_{HS}}\right)^2}+\frac{c_1}{1+\frac{c_2 R}{R_{HS}}}\ ,
\label{3.7}
\ee
where we have assumed $n=1$ in (\ref{1.8a}), as in the previous section. Then,  the scalar potential (\ref{3.3}) yields,
\be
V_{HS}(\phi)=\frac{1+c_1-\phi\pm2\ c_2\sqrt{c_1\ (1-\phi)}}{c_2}R_{HS}\ , 
\label{3.8}
\ee
Hence, the scalar-tensor representation is not uniquely defined, but the potentials exhibit two branches, which in principle do not affect the cosmological evolution (see Fig.~\ref{fig3}), but it may influence the behavior of the phase space. In addition, the model introduces a boundary condition on the value of the scalar field, being $\phi<1$, a limit where both branches of the potentials converge, and where the first derivative of the potential exhibits a pole, which induces a sudden singularity (see Ref.~\cite{Appleby:2009uf}). Note that there is a direct correlation between the behavior of the scalar field and the evolution of the EoS parameter as can be shown comparing figs.~\ref{fig1} and \ref{fig1a} with fig.~\ref{fig3}: the EoS parameter presents some oscillations, specially for negative redshifts around the phantom barrier, while the evolution of the scalar field also oscillates at negative redshifts. 

Let us now analyze the phase space $\{H,\phi\}$ for this model in vacuum. For simplicity, we assume vacuum FLRW equations, so that by combining FLRW equations (in the scalar-tensor equivalence), the following equations are obtained,
\be
\dot{H}=-2H^2+\frac{1}{6}V'(\phi)\ , \quad \dot{\phi}=\frac{1}{3H}\left[-3H^2\phi+\frac{1}{2}V(\phi)\right]\ .
\label{3.9}
\ee
which yield,
\be
\frac{dH}{d\phi}=\frac{\frac{V'(\phi)}{6}-2H^2}{\frac{V(\phi)}{3H}-H\phi}\ .
\label{3.10}
\ee
\begin{figure}[h!]
  \resizebox{14pc}{!}{\includegraphics{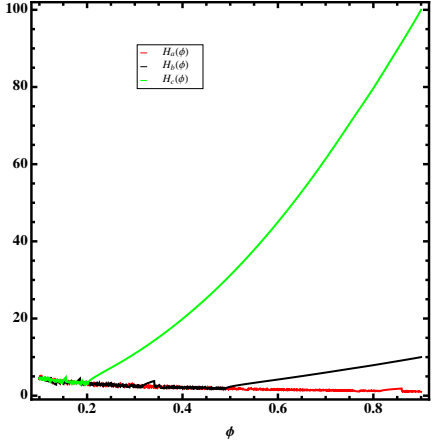}}
  %\resizebox{13.7pc}{!}{\includegraphics{1b}}
\caption{Phase space $H(\phi)$ for the HS model, where $V_{+}$ in (\ref{3.8}) is considered. We have assumed initial conditions close to the boundary of the scalar field: $(a)\  H(0.95)=H_0^2$, $(b)\  H(0.95)=10H_0^2$, $(c)\ H(0.95)=100H_0^2$.}
\label{fig4}
\end{figure}
This equation describes the phase space for a particular scalar potential. Fig.~\ref{fig4} the phase space for the model (\ref{1.8a}) with the same free parameters assumed above. Then, it shows that regardless of the choice of  initial conditions, the behavior of the Hubble parameter tends to a stable point, close to zero, where some tiny fluctuations are observed. 
Other viable models have a similar behavior, where tiny differences between the models are observed when the future evolution of the EoS is studied, as pointed out in Ref.~\cite{SaezGomez:2012ek}. 

\section{Discussions}
Here, we have analyzed the future evolution of a particular viable $f(R)$ gravity model, \cite{f(R)viable1}. The cosmological evolution of the effective EoS enters into a phantom stage in the present or near future, while the cosmological evolution, similarly to $\Lambda$CDM model at $z=0$, is well reproduced for small redshifts, where we have imposed some particular conditions in order to fit the model with the observational constraints. On the other hand, the transition to the phantom phase does not imply directly the occurrence of a future singularity, since the Hubble parameter and its first derivatives remain finite. In addition, the scalar-tensor counterpart of the $f(R)$ model shows a well behavior of the phase space, at least when $\phi<1$, which is the case for small redshifts as shown in Fig.~\ref{fig3}.  Nevertheless, this does not prevent that a cosmological singularity, induced by the pole of the scalar potential, may occur in the past. However, for large redshifts, one expects that modifications may be negligible, so that the singularity may be avoided. A deeper study on this issue should be done in future works.  
Moreover, the possibility of a {\it Little Rip}, a phase where the strength of the expansion would be comparable with some bounded systems as the Solar system (see Ref.~\cite{Frampton:2011rh}), seems to be far to occur in viable $f(R)$ gravity, as pointed out in Ref.~\cite{SaezGomez:2012ek}. Hence, viable modified gravities have still a strong support to be considered as a serious candidate to dark energy.

\begin{theacknowledgments}
 I acknowledge support from a postdoctoral contract from the University of the Basque Country (UPV/EHU) under the program ``Specialization of research staff'', and support from the research project FIS2010-15640, and also by the Basque Government through the special research action KATEA and UPV/EHU under program UFI 11/55.
\end{theacknowledgments}

% choose bibtex style depending on layout style and options used in
% sample:

%\doingARLO[\bibliographystyle{aipproc}]
%          {\ifthenelse{\equal{\AIPcitestyleselect}{num}}
%             {\bibliographystyle{arlonum}}
%             {\bibliographystyle{arlobib}}
%          }
\bibliographystyle{aipproc}   % if natbib is available

\end{document}